# Rentable Internet of Things Infrastructure for Sensing as a Service (S2aaS)

Charith Perera

Sensing as a Service ($S^2$aaS) model [1] [2] is inspired by the traditional Everything as a service (XaaS) approaches [3]. It aims to better utilize the existing Internet of Things (IoT) infrastructure. $S^2$aaS vision aims to create '*rentable infrastructure*' where interested parties can gather IoT data by paying a fee for the infrastructure owners.

Read More

$S^2$aaS model primarily utilizes the existing IoT infrastructure which is being deployed to achieve a primary objective. For example:

- a shop may deploy a security camera system in order to provide security for its premises (*primary objective*). However, such cameras (or the data captured by the cameras) can be reutilized (or reanalyzed) to understand the consumer patterns (e.g., analyze demographics such as age, gender, etc. of the people who pass by).
- a garbage bin may be fitted with sensors in order to monitor and track garbage levels and to support resource management (e.g., truck allocation, recycling facility demand monitoring etc.). Same sensing infrastructure can also be reutilized to understand crowd in a given day (e.g., understand crowds based on what they throw away).

Let us consider the following scenario, as illustrated in Figure 1. There is a game in the stadium on the weekend. A marketing company, *BestBrands*, wants to understand the attending crowds better to develop their promotional campaigns specifically targeting the spectators (market segment). Therefore, they may be interested in collecting data such as demographics (age ranges, gender, sentiments, etc), movement, sentiments, buying behaviors, etc. Through a broker, *BestBrands* aims to rent the infrastructure over a certain period of time (during the game day), so they can gather the data in order to understand the crowd better. *BestBrands* may be interested to gather a variety of data from the streets. Different sensors may be used to gather and infer different types of knowledge: video cameras [demographics]; motion sensors: [number counting, crowd movement identification]; environmental sensors (e.g. temperature, wind, humidity): [identify any influencing factors, buying behaviors, etc].

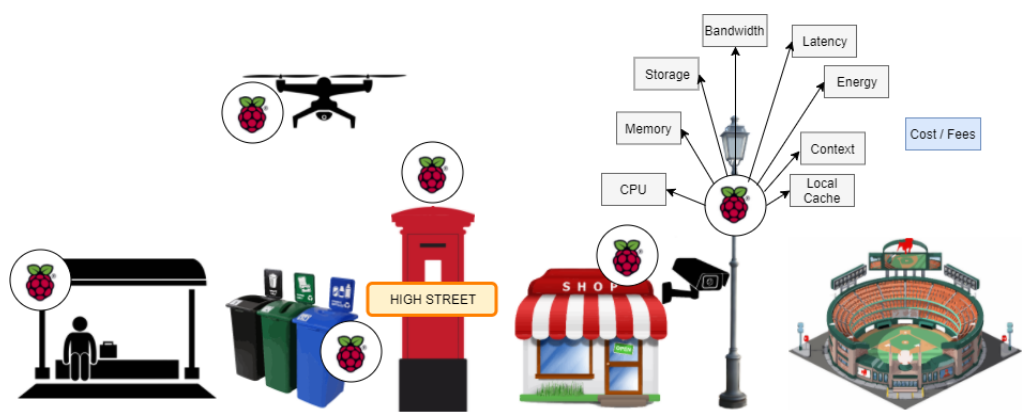

**Figure 1: Cloud Initiated Sensing as a Service.**

**Major Research Challenges**

There are many research challenges that need to be addressed in order to realize this vision. Optimum IoT resources orchestration to facilitate users' requirements is one of the major challenges. Such orchestrations should also respect user preferences and while managing the overall efficiency of the network. In the above context, *BestBrands* may either interest in gathering data in real-time (e.g., to enrich their promotion in real-time) or in a differed manner (e.g., to enrich future promotional campaigns). The orchestrations need to be performed accordingly to support the two types of sensing requirements. In order to support real-time sensing as a service, orchestration will be required to bring more computational nodes together in order to process data at a higher rate to reduce latency. Due to high resource consumption (both computation and network), *BestBrands* will be required to pay a higher price. Knowledge engineering techniques (e.g., semantic web technologies) can be used to enable the optimum IoT resource orchestration process in conjunction with AI planning techniques [4].

One of the major challenges in edge computing is to reduce network communication and latency. *Knowledge engineering techniques* can be used to enrich edge nodes with intelligence (knowledge), so they can make decisions by themselves reducing communication with the cloud. Orchestration also requires discovering IoT resource (e.g. computational nodes, service, sensing capabilities, etc.) efficiently in order to develop an optimal plan at runtime. Knowledge engineering techniques are also useful towards performing ad-hoc resource discovery.

In the above use case, the orchestration is triggered via a cloud broker where the *BestBrands* makes its initial request. However, there is another type of scenarios that could occur as follows where the request initiated by one of the edge nodes. Let us consider the scenario presented in Figure 2.
Bob is visiting a tourist attraction and he is interested in using his augmented reality device (AR) (mobile phone, glasses, etc.) to enrich his experience. He is interested in a rich experience, so he would like to rent nearby IoT infrastructure to support the experience. His augmented reality device would discover the nearby infrastructure to share the computation load (computation offloading), so Bob's own AR device can reduce its energy consumption. As a result, Bob can have longer experience. Bob's AR device will orchestrate the different computational tasks to different nodes (e.g., download and process maps, weather information, audio narration, translation, etc.). Such distribution of tasks will reduce the latency and improve Bob's experience. Bob is happy to pay for this rich experience.
On the other hand, Alice is a university student with a limited budget. She is less concerned about the experience, but she needs to retain the battery of the mobile phone until she returns back to the hotel. Based on her priority, the orchestration that Alice's AR device need to perform would be significantly different from Bob's orchestration. Alice may pay less than Bob, but her experience may not as rich as Bob's (e.g., latency, feature limitations).

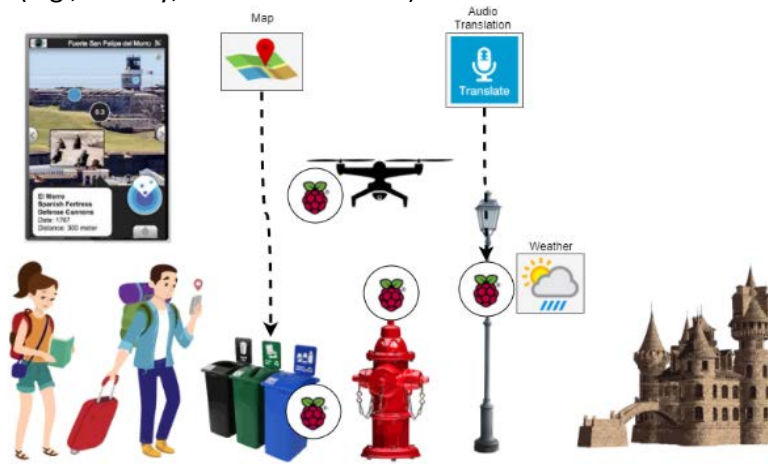

**Figure 2: Edge Initiated Sensing as a Service.**

In this scenario, the request is initiated by Alice's and Bob's AR devices (edge devices). As same as in the previous scenario, orchestration may need to consider contextual information. Candidate compute nodes may not only have different computational and sensing capabilities, but they may also have other relevant resources already with them. For example, the garbage bin may already have the map in its local cache (that both Alice and Bob needs). Therefore, it is much efficient to assign map processing to the garbage bin node. Similarly, there could be many considerations that the orchestration algorithms need to consider (in addition to user preferences). *Knowledge engineering techniques* (interoperability, semantics) can play a significant role in edge orchestration activities. Even though service composition for the ubiquitous domain is well researched (though mostly in simulations), they all assume nodes and the services are inseparable and static [5].

In contrast, one of the main assumptions in S$^2$aaS is that infrastructure and associated resources are rentable, and the services are separable from nodes. This means that the assignment of services into rented compute nodes happens dynamically. Such separability allows performing orchestration in a much fine-grained and optimum manner. However, such separability also makes discovery and orchestrating algorithms much more complex (due to increased possibilities) than typical service composition. Therefore, new algorithms will be required to tackle this challenge efficiently.

In additional to the rentable infrastructure already deployed across cities, we envision that some service provider may deploy purpose build devices (e.g., drones augmented with rentable infrastructure) in high demand areas. It is also interesting to exploring how such services can be commissioned in real-world scenarios.

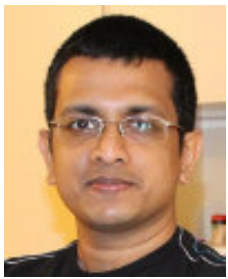

**Charith Perera** is a Research Associate at Newcastle University, UK. He received his BSc (Hons) in Computer Science from Staffordshire University, UK and MBA in Business Administration from the University of Wales, Cardiff, UK and Ph.D. in Computer Science at The Australian National University, Canberra, Australia. Previously, he worked at the Information Engineering Laboratory, ICT Centre, CSIRO. His research interests are Internet of Things, Sensing as a Service, Privacy, Middleware Platforms, and Sensing Infrastructure. He is a member of both IEEE and ACM. Contact him at www.charithperera.net or charith.perera@ieee.org